\documentclass[usenatbib]{mn2e}
\usepackage[totalwidth=170mm, totalheight=240mm, paperwidth=8.5in,  paperheight=11in]{geometry}

\usepackage{amsfonts}
\usepackage{amsmath,amssymb}
\usepackage{graphicx}

\newcommand{\lb}[0] { \left( }
\newcommand{\rb}[0] { \right) }
\newcommand{\beqs} { \begin{eqnarray} }
\newcommand{\eeqs} { \end{eqnarray} }
\newcommand{\bsub} { \begin{subequations} }
\newcommand{\esub} { \end{subequations} }

\newcommand{\degree}{\ensuremath{^\circ}}

\newcommand{\pI}{paper I}
\newcommand{\pIsp}{paper I }

\begin{document}


\title{Note on galaxy catalogues in UHECR flux modelling}

\author[Hylke Koers and Peter Tinyakov]
{Hylke B. J. Koers$^{1}$\thanks{E-mail: hkoers@ulb.ac.be} and
Peter Tinyakov$^{1,2}$\thanks{E-mail: Petr.Tiniakov@ulb.ac.be} \\
$^{1}$Service de Physique Th\'eorique, Universit\'e Libre de Bruxelles (U.L.B.), CP225, Bld. du Triomphe, B-1050 Bruxelles, Belgium\\
$^{2}$Institute for Nuclear Research, 60th October Anniversary Prospect 7a, 117312, Moscow, Russia}

\pagerange{\pageref{firstpage}--\pageref{lastpage}} \pubyear{2009}
\maketitle
\label{firstpage}

\begin{abstract}
We consider the dependence of ultra-high energy cosmic ray (UHECR) flux predictions on the choice of galaxy
catalogue. We demonstrate that model predictions by \citet{Koers:2008ba},
based on the so-called KKKST catalogue, are in good agreement with
predictions based on the XSCz catalogue,
a recently compiled catalogue that contains spectroscopic redshifts
for a large fraction of galaxies.
This agreement refutes the
claim by \citet{Kashti:2009ui} that the KKKST catalogue is not suited
for studies of UHECR anisotropy
due to its dependence on photometric redshift estimates.
In order to quantify the effect of galaxy catalogues on flux predictions,
we develop a measure of anisotropies associated with model flux maps.
This measure offers a general criterion to study the effect of
model parameters and assumptions
on the predicted strength of UHECR anisotropies.

\end{abstract}


\section{Introduction}
Galaxy catalogues are an
indispensable tool to model the structure of the local Universe,
as required for studies of ultra-high energy cosmic ray (UHECR) anisotropy.
Many such studies (e.g. \citealt{Cuoco:2005yd,  Kashti:2008bw,  Takami:2008ri, Kashti:2009ui})
have used the PSCz catalogue \citep{Saunders:2000af}
for this purpose. This catalogue, however, has its drawbacks: it suffers from incomplete
sky coverage, it may underestimate galaxy counts
in high-density regions \citep{HuchraWebsite}, and it has limited statistics.
The 2MASS galaxy catalogue \citep{Jarrett:2000me,2004PASA...21..396J, 2006AJ....131.1163S},
offering complete sky coverage
(except for the galactic plane) and excellent statistics, improves on these issues.
However, this catalogue does not contain
spectroscopic redshifts so that redshifts have to be estimated by
photometry, i.e. from the observed brightness under the assumption
of a standard intrinsic brightness in a specific
band. Efforts to
obtain spectroscopic redshifts for 2MASS  galaxies
are ongoing. The resulting catalogue, termed XSCz, is 
presently being compiled.\footnote{See \texttt{http://web.ipac.caltech.edu/staff/jarrett/XSCz/}.}

Motivated by the imperfection of existing galaxy catalogues, \citet{Kalashev:2007ph}
have compiled a hybrid catalogue that contains galaxies with spectroscopic redshifts from
the (HYPER)LEDA database \citep{2003A&A...412...45P} within 30 Mpc
and galaxies from the 2MASS Extended Source catalogue \citep{Jarrett:2000me}
at larger distances, for which photometric redshift estimates
are used (the two contributions are of course normalized
appropriately). This catalogue, which we have termed the KKKST catalogue,
is used in \citeauthor{Koers:2008ba} (\citeyear{Koers:2008ba}; hereafter \pI)

The adequacy of the KKKST catalogue for UHECR flux
modelling is questioned by \citet{Kashti:2009ui}  because of its dependence
on photometric redshift estimates. The errors in these estimates give rise to errors in flux
estimates, which may distort model flux maps and may lead to inaccurate model predictions.
The uncertainty in photometric redshift estimates is indeed
fairly large: \citet{Jarrett:2004nk} quotes
an error of $10\%-20\%$ for most normal galaxies in the 2MASS survey
(\citet{Kashti:2009ui} claims a $\sim$$30\%$ systematic uncertainty).
In \pI, we deemed these uncertainties acceptable for two
reasons.
First, large-scale anisotropies in flux maps arise as a collective
effect of many sources. Adding flux contributions of many
individual galaxies reduces the relative strength of fluctuations
(for the KKKST catalogue with a smearing angle of
$\theta_{\rm s}=6\degree$, as adopted in \pI,
a model flux is composed of individual contributions of
$\sim$$600$ galaxies).
Second, the remaining uncertainties only
affect the model flux beyond 30 Mpc, while the imprint of local
structure is strongest at close distances.

In this paper we compare model fluxes based on the KKKST catalogue
to model fluxes based on a preliminary version of the XSCz
catalogue.\footnote{We thank Tom Jarrett for providing us with a preliminary version
of the XSCz catalogue.}
This allows us to
determine to which extent the flux predictions in \pIsp are
contaminated by the inaccuracies in the catalogue used.
The XSCz catalogue is very well suited to model the matter distribution
in the Universe because of its completeness, large statistics, and
spectroscopic redshift measurements.
We therefore
consider the resulting model fluxes as a benchmark against which 
UHECR flux predictions derived with the KKKST catalogue (as well as with the
PSCz catalogue) can be cross checked. 
As we will demonstrate, flux predictions based on the KKKST catalogue (as well as on
the PSCz) are in good agreement with the XSCz prediction.
This provides an \emph{a posteriori} verification
of the accuracy of our models in \pI.

We also investigate, in a general setup, the relationship between
galaxy catalogues and the predicted strength of UHECR anisotropies.
For this purpose we develop a measure
that quantifies the relation between model flux maps, which represent the model predictions for a given galaxy catalogue,
and the test statistic $D$ (introduced in \pI), which
is a measure of the strength of UHECR anisotropies.
This measure can be used to assess the predicted
strength of anisotropies from the ``contrast'' that is exhibited in
model flux maps and is independent of event number.
Because of its generality, this measure may also be useful
to study the effect of other
model parameters and assumptions (e.g., threshold energy, deflection
angle, or injection spectrum)
on the predicted strength of UHECR anisotropies.

This paper is organized as follows.
In section \ref{section:KKKST} we compare model predictions from the KKKST
and PSCz catalogues to predictions based on the XSCz catalogue.
Section \ref{section:fluxmapcontrasts} concerns
the relationship between
galaxy catalogues and the predicted strength of UHECR anisotropies.
We summarize our findings in 
section \ref{sec:summary}.

\section{Accuracy of model flux maps}
\label{section:KKKST}

\begin{figure}
\begin{center}
\includegraphics[width=5.5cm, angle=270]{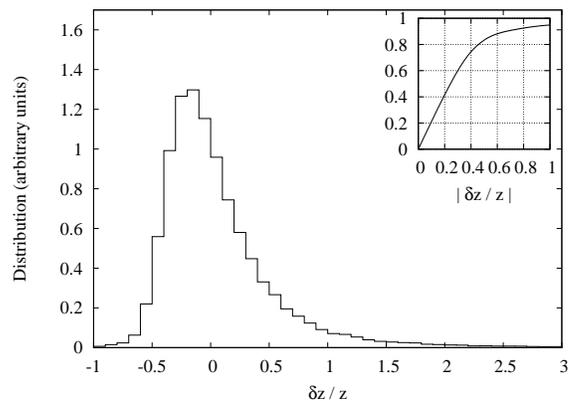}
\caption{\label{fig:diffz} Distribution of the relative difference 
$\delta z/z$ between
photometric and spectroscopic redshift estimates for individual
galaxies in the XSCz catalogue. Inset:
cumulative distribution of $|\delta z/z|$. In producing
this figure we only used galaxies with $D>5$ Mpc.}
\end{center}
\end{figure}

We begin our analysis by considering the accuracy of photometric
redshift determinations in the 2MASS catalogue.
We compute
the relative difference \mbox{$\delta z/z := (z_{\rm photo}-z_{\rm spec})/z_{\rm spec}$},
where $z_{\rm photo}$ denotes a photometric redshift estimate
and $z_{\rm spec}$ denotes a spectroscopic redshift,
for all galaxies in the XSCz catalogue
that have spectroscopic redshifts.
Here, and throughout the paper, we 
only consider galaxies with $K_s$-magnitude below 12.50 in the XSCz catalogue.
Spectroscopic redshifts are available
for $\sim$70\% of these relatively bright galaxies.
The
distribution of $\delta z/z$ is shown in Fig.~\ref{fig:diffz}.
The inaccuracy in photometric redshift estimates
is indeed fairly large: we find that the average value of $|\delta z/z|$, the absolute relative difference
in redshift determinations, is 0.36.


\begin{figure}
\begin{center}
\includegraphics[width=5.5cm, angle=270]{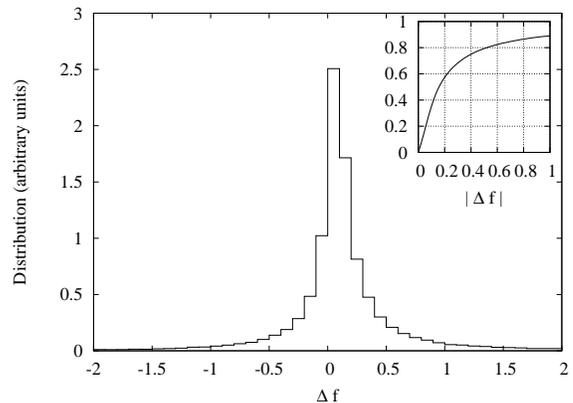}
\caption{\label{fig:diffflux_singlesource} Distribution of 
$\Delta f$ for individual
galaxies in the XSCz catalogue. In producing this plot, 
we approximated fluxes as $\phi \propto D^{-2}$ and
we only considered galaxies with $D>5$ Mpc.}
\end{center}
\end{figure}

The UHECR luminosity at the source being unknown,
model flux predictions are conveniently formulated in
terms of the normalized flux
\beqs
f = \frac{\phi}{\langle \phi \rangle} \,,
\eeqs
where $\phi$ denotes a single-source flux and
$\langle \phi \rangle$ is the average flux of all galaxies.
(Note that the overall flux normalization
is determined by observations).
To get a rough estimate of the inaccuracies in flux predictions at the level
of individual sources, we 
approximate the flux associated with every galaxy as $\phi \propto D^{-2}$,
$D$ being the source distance (we take $D \propto z$ here).
Here, and throughout this paper,
we assume all sources have the same intrinsic luminosity.
We then compute the quantity $\Delta f = f_{\rm photo} - f_{\rm spec}$, 
where $f_{\rm photo}$ is the normalized flux estimate based on the photometric
redshift estimate and $f_{\rm spec}$ is the normalized flux  derived from
the spectroscopic redshift. The distribution of 
$\Delta f$ is shown in Fig.~\ref{fig:diffflux_singlesource}.



\begin{figure}
\begin{center}
\includegraphics[width=7cm]{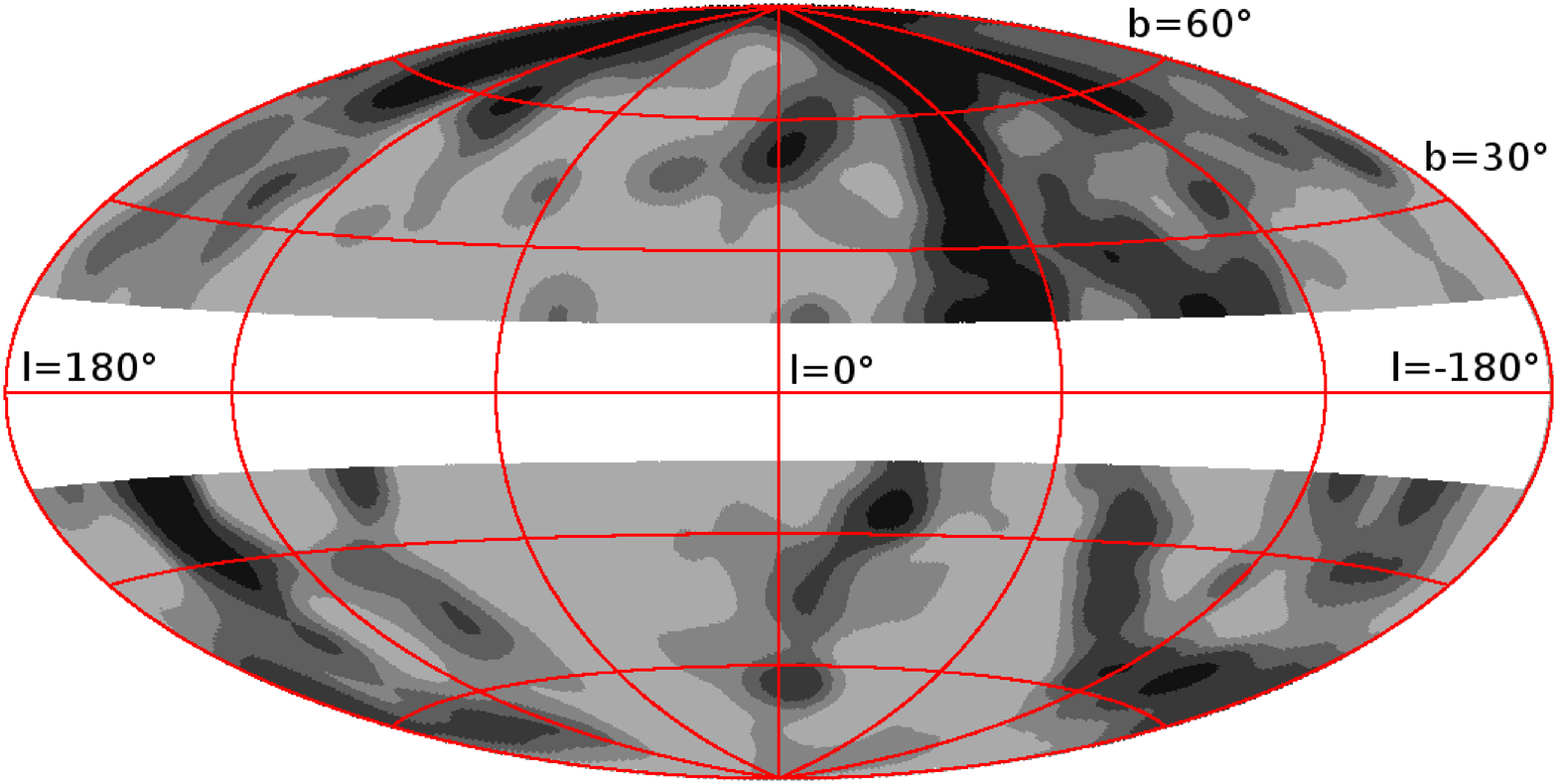}
\vspace{0.1cm}

\includegraphics[width=7cm]{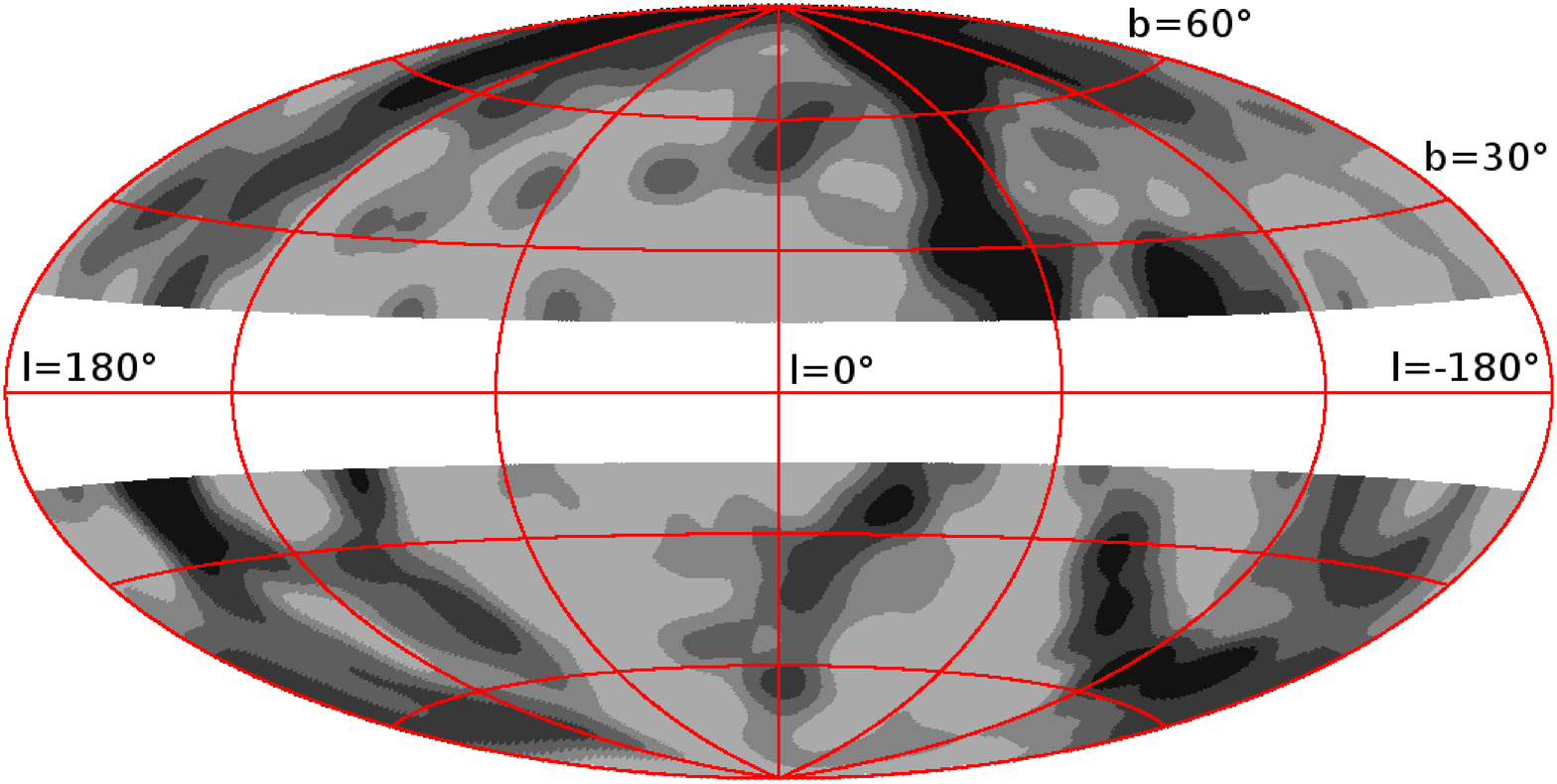}
\vspace{0.1cm}

\includegraphics[width=7cm]{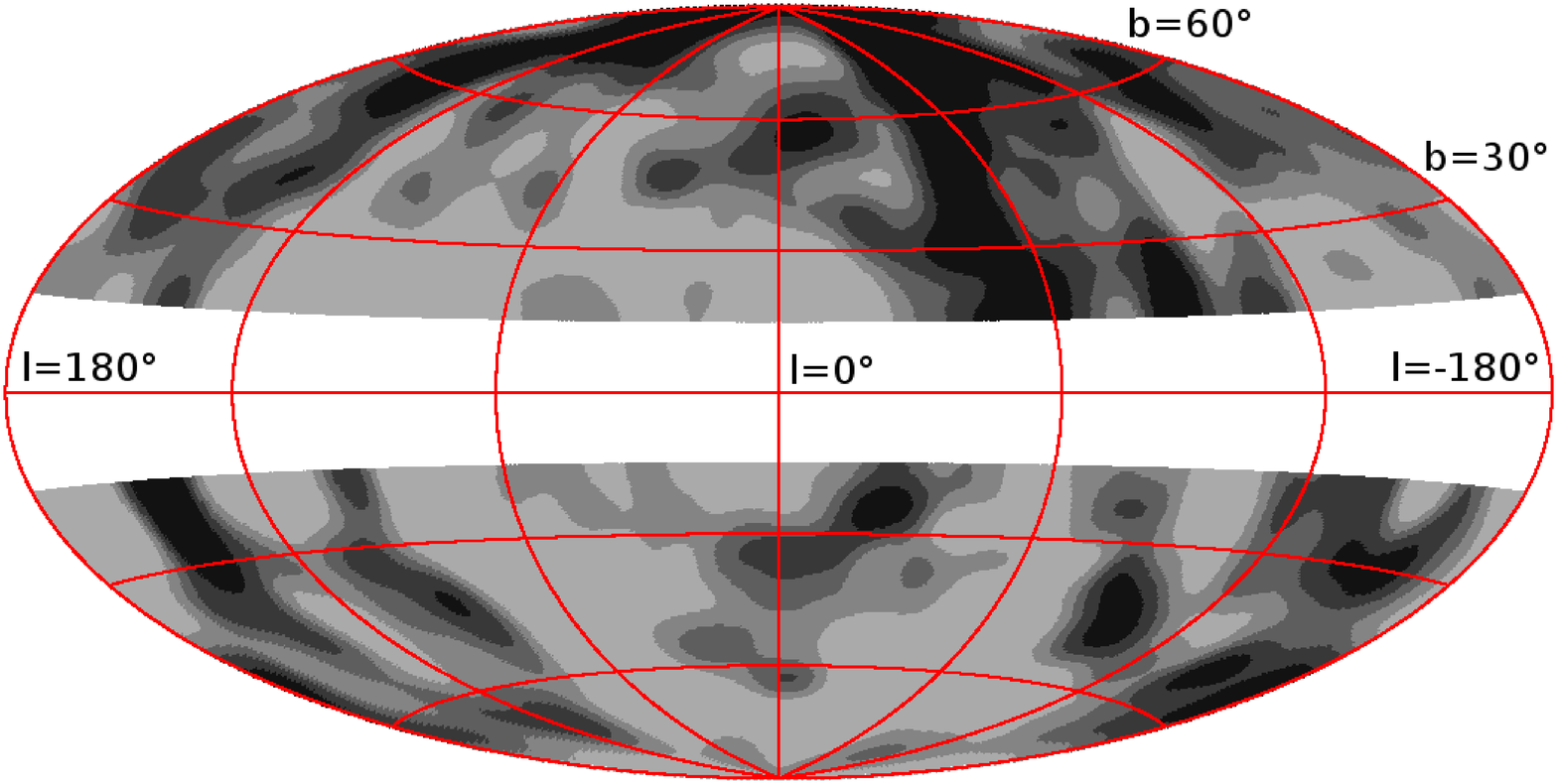}
\caption{\label{fig:skymaps} UHECR model flux maps
for  $E_0=60$ EeV, $\theta_{\rm s}=6\degree$, and uniform exposure.
The three panels show model predictions
based on different galaxy catalogues:
the KKKST catalogue (top panel), the XSCz catalogue (middle) and 
the PSCz catalogue (bottom). The gray bands
are chosen such that each contains 1/5 of the model flux,
with darker gray indicating larger flux. }
\end{center}
\end{figure}

\begin{figure}
\begin{center}
\includegraphics[width=5.5cm, angle=270]{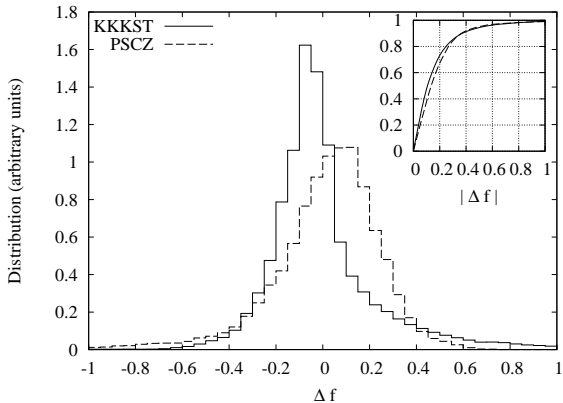}
\caption{\label{fig:diffflux} 
Distribution of $\Delta f$ for the model fluxes 
shown in Fig.~\ref{fig:skymaps}.}
\end{center}
\end{figure}

We now consider model UHECR flux maps, which are generated using the methods
described in \pI. 
Throughout this work we assume a proton composition and a
power-law injection spectrum with spectral index $p=2.2$
extending to  very high  energies.
We adopt a $\Lambda$CDM concordance model with Hubble
constant \mbox{$72$ km s$^{-1}$ Mpc$^{-1}$}, and cosmological density parameters
$\Omega_m = 0.27$ and  $\Omega_{\Lambda} = 0.73$.
Following the setup in \pI, we adopt a smearing angle
$\theta_{\rm s}=6\degree$. Using the method described in \cite{Koers:2009pd},
we have verified that all catalogues contain sufficient galaxies
to provide a good statistical description on this angular scale.
We remove the region $|b|<15\degree$ from our analysis because of incompleteness near
the galactic plane. (We choose the same region for every catalogue for comparison; for the XSCz skymap
this choice is overly conservative).
Fig.~\ref{fig:skymaps} shows the flux maps
for a threshold energy $E_0 = 60$ EeV. 
In Fig.~\ref{fig:diffflux} we show the distribution of flux differences
$\Delta f = f_{\rm alt} - f_{\rm XSCz}$, where
$f = \phi / \langle \phi \rangle$ denotes the normalized flux and the subscript refers to the catalogue
that was used in the modelling (``alt'' standing for either KKKST or PSCz).
Here $\phi$ represents the integral UHECR flux in a given direction
(as represented in Fig.~\ref{fig:skymaps})
and $\langle \phi \rangle$ is the average flux on the sphere.

Figures~\ref{fig:skymaps} and \ref{fig:diffflux} demonstrate
that model predictions based on both the KKKST and PSCz catalogues are
in good agreement with those from the XSCz catalogue. 
The average value of $|\Delta f|$ is 0.17 for the KKKST catalogue
and 0.19 for the PSCz.  At lower threshold energies the differences
become smaller, while at higher energies they
are somewhat larger. At $E_0=100$ EeV, for example, the average
$|\Delta f|$ is 0.39 for the KKKST catalogue
and 0.32 for the PSCz.  

Sampling a flux map uniformly over the sky, one obtains a flux distribution associated with the map.
This distribution encodes the relevant intrinsic properties of the flux map, i.e. properties relating to
the strength of over- and underdense regions but not to their position on the sky.
Postponing a more thorough discussion of the flux distribution to the next section,
we point out here that the flux distribution
corresponding to the KKKST flux map 
is broader than the flux distribution obtained with the 
PSCz catalogue (the XSCz is in between). 
This is reflected in the fact that the band of highest flux
(darkest gray) in Fig.~\ref{fig:skymaps} occupies the smallest
area on the sphere: 7.5\% compared to 8.2\% for the XSCz flux map
and 10.4\% for the PSCz flux map. For an isotropic flux map this number
would be $1/5=20$\%. We see that the KKKST map deviates most
strongly from an isotropic flux map. As a consequence, model predictions based on
this catalogue will exhibit the strongest departures from isotropy. We will use the term
``contrast'' to refer to the width of the flux distribution, strong contrast corresponding
to a wide flux distribution and a strong deviation from isotropy.

\begin{figure}
\begin{center}
\includegraphics[width=5.5cm, angle=270]{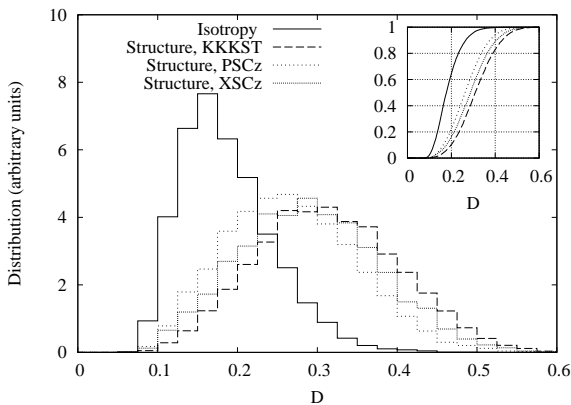}
\caption{\label{fig:dist-D} Distribution of test statistic $D$
for an isotropic reference distribution and test UHECR event sets
following the distribution of matter as modelled from different
catalogues. This figure applies to 20 events with threshold energy 60 EeV, and the
distributions are obtained by sampling over
$10^4$ independent Monte Carlo realizations.}
\end{center}
\end{figure}

How does the choice of galaxy catalogue affect the statistical tests 
proposed in \pI? We consider here the $D$-test, to
which we will refer as the ``flux sampling method''.
The test defines a test statistic  $D$ as the Kolmogorov-Smirnov (KS) distance $D_{\rm KS}$ between
the cumulative distribution of flux values sampled by a set of UHECR events
and a reference distribution that corresponds to the model that is tested
(see \pIsp for details and discussion). Here and in the following we
 consider two models: the ``Isotropy'' model (denoted  $\mathbf{I}$), which states
that UHECR events are distributed isotropically (we do not consider experimental exposure here), and the
``Structure'' model (denoted  $\mathbf{S}$), which states that UHECR sources trace the distribution
of matter in the Universe.
As a case study, we show in Fig.~\ref{fig:dist-D} the distribution
of the test statistic $D$ for 20 UHECR events with energies in excess of 60 EeV. 
Here the reference distribution corresponds to an isotropic
flux, so that the curve labeled ``Isotropy'' follows the universal
KS self-correlation distribution (it is thus the same for all catalogues).
The ``Structure''  distribution obtained from the PSCz catalogue
is clearly closer to the ``Isotropy'' distribution than the
``Structure''  distribution obtained from the KKKST catalogue
(the distribution from the XSCz being intermediary).
This is reflected in the
statistical power estimates:
For a significance $\alpha=0.05$, we find $P = 0.54$ for the
KKKST catalogue, $P=0.46$ for the XSCz catalogue, and
$P=0.37$ for the PSCz catalogue, where $P$ denotes the power to
reject $\mathbf{I}$ when $\mathbf{S}$ is true.
The difference in statistical powers can be related to the contrast in the model flux maps
discussed in the previous paragraph.
This relation will be explored in the
next section.

We have recomputed power estimates and $p$-values for the $D$-test
reported in \pIsp with the XSCz catalogue instead 
of the KKKST catalogue. At energies below 60 EeV we find that the power
estimates change by only a few percent. At 100 EeV the XSCz catalogue leads 
to power estimates smaller by $\sim$20\%, which is still acceptable
in the light of other uncertainties. For the data obtained by the Pierre
Auger Observatory and the AGASA experiment, we find no significant 
difference in $p$-values for the different catalogues.

\section{Galaxy catalogues and predicted strength of anisotropies}
\label{section:fluxmapcontrasts}
In this section we investigate how the choice of catalogue affects the predicted strength of
large-scale UHECR anisotropies. In order to quantify this strength we use the
flux sampling test to measure deviations from isotropy. Although other tests will measure
anisotropies in a different manner, a galaxy catalogue that yields a value for the $D$ test statistic
close to the isotropic prediction will in general also yield outcomes close to the isotropic
prediction in other tests (and similarly for predictions far away from the isotropic one).

The outcome of the flux sampling test depends on the number of events  as well as on
the distribution of model fluxes over the sky, i.e. the flux map. The first parameter is obviously
independent of the choice of galaxy catalogue. All intrinsic properties of a galaxy catalogue that may
affect the strength of anisotropies are contained in the flux map. Here and in the following, 
the expression ``flux map'' is used to refer to the flux map derived under $\mathbf{S}$ 
(i.e., assuming UHECR sources trace the distribution of matter in the Universe).
The flux map also depends on threshold energy, UHECR injection spectrum,
and average deflection angle.


\begin{figure}
\begin{center}
\includegraphics[width=5.5cm, angle=270]{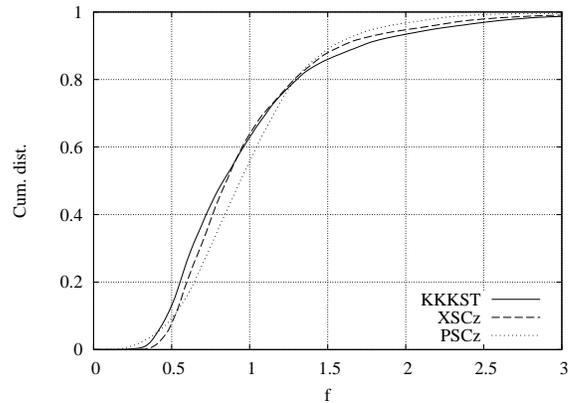}
\caption{\label{fig:intrinsicflux}  The cumulative distribution
$\mathcal{C}_{\rm iso} (f)$
for the KKKST, PSCz, and XSCz catalogues. This figure applies to
threshold energy $E_0=60$ EeV and smearing angle $\theta_{\rm s}=6\degree$.}
\end{center}
\end{figure}

The connection
between flux maps and the strength of UHECR anisotropies can be illustrated
by two limiting cases. First, consider the case that $\mathbf{S}$ 
would predict a uniform flux, so that the flux distribution would be a delta-function (minimal contrast).
In that case UHECR events
sample the sky uniformly under both $\mathbf{S}$  and $\mathbf{I}$ 
and model predictions become identical. In the opposite case of
a very wide flux distribution (strong contrast),
events will have  a strong tendency to cluster in high-flux regions
under $\mathbf{S}$, exhibiting strong anisotropy.
The connection between flux distributions and the strength of anisotropies can be made
explicit using the $D$-test. As shown in 
appendix \ref{app:DKS},  under $\mathbf{S}$ the test statistic $D$ approaches the following
limiting value as the number of events goes to infinity
(recall that it approaches 0 in the same limit under $\mathbf{I}$):
\beqs
\label{eq:DKS-lim} D  \to \int_0^1 d f \, \mathcal{C}_{\rm iso} (f) \, ,
\eeqs
where $\mathcal{C}_{\rm iso} (f)$ denotes the normalized
cumulative flux distribution under $\mathbf{S}$
and we recall that $f = \phi/\langle \phi \rangle$, $\phi$ being
the integral UHECR flux and $\langle \phi \rangle$ the average value on
the sphere.
Equation \eqref{eq:DKS-lim} identifies the intrinsic property of a flux map that is 
important in determining the strength of UHECR anisotropies.
It gives a precise meaning to the concept of ``contrast'' that was
introduced in the previous section. It is related to the width of the
flux distribution: A narrow flux distribution corresponds to a steeply rising cumulative
distribution and, via equation~\eqref{eq:DKS-lim}, a value of $D$ close to 0.


\begin{figure}
\begin{center}
\includegraphics[width=5.5cm, angle=270]{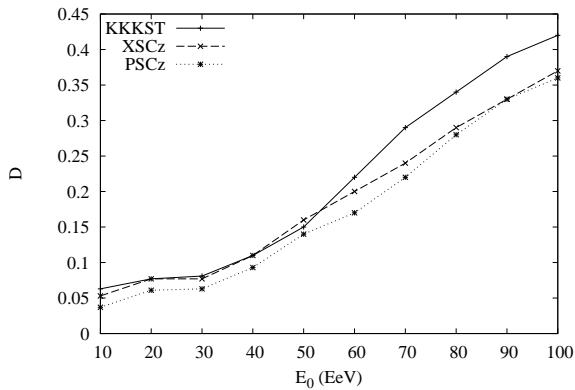}
\caption{\label{fig:DKS-E} Asymptotic value of
$D$ for $N_{\rm ev} \to \infty$ for the KKKST, XSCz, and PSCz catalogues
as a function of $E_0$.}
\end{center}
\end{figure}

The distribution $\mathcal{C}_{\rm iso}$ is shown in 
Fig.~\ref{fig:intrinsicflux} for the three flux maps presented in 
Fig.~\ref{fig:skymaps}. Comparing the surfaces under the curves
for $0<f<1$, it is clear that the PSCz catalogue yields the smallest
asymptotic value of $D$, and the KKKST catalogue the largest. 
The larger contrast in the KKKST catalogue implies stronger
signatures of anisotropy. This explains
the ordering of statistical powers that was found in the
previous section.

In Fig.~\ref{fig:DKS-E} we show the 
asymptotic value of $D$ from eq.~\eqref{eq:DKS-lim} for
different catalogues and threshold energies.
We observe that the KKKST catalogue systematically predicts somewhat
stronger anisotropies compared to the PSCz catalogue, in keeping with the
results presented above and with the discussion in
\pI.\footnote{Unfortunately, the power estimates
for the PSCz catalogue as reported in \pIsp are inaccurate
due to an erroneous application of the PSCz selection
function in our numerical routines. In table 3 the power estimates for
the PSCz catalogue should read
0.37, 0.28, and 0.72 for scenarios I, II, and III, respectively.
With these changes,
the statistical powers for the PSCz catalogue remain smaller
than those for the KKKST catalogue so that
the conclusions remain qualitatively unchanged.}
The XSCz prediction lies
in between. For threshold energies $E_0 \lesssim 60$ EeV, 
the XSCz curve is virtually identical to the KKKST one, while it approaches the PSCz curve
at higher energies. 
The differences between the three curves are however small. In particular,
they are smaller than the uncertainty
induced by systematic errors in energy determination. (One may check that
the curves corresponding to the KKKST and PSCz catalogues are within
a 20\% shift in energy applied to the XSCz prediction).

\begin{figure}
\begin{center}
\includegraphics[width=5.5cm, angle=270]{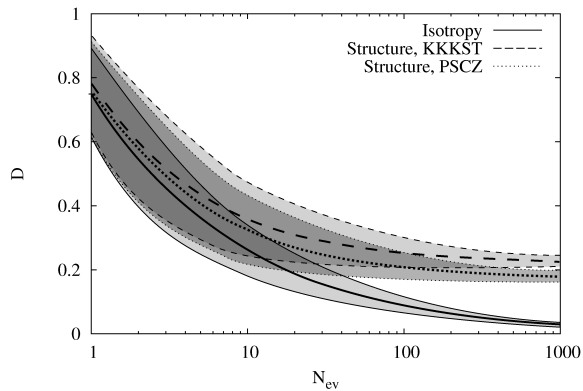}
\caption{\label{fig:DKS-NeV} Average value of
$D$ (thick lines) and 1-sigma contours (thin) as a
function of event number $N_{\rm ev}$ for the KKKST and PSCz
catalogues and for $E_0$ = 60 EeV. The full distribution
for $N_{\rm ev}=20$ is shown in
Fig.~\ref{fig:dist-D}.}
\end{center}
\end{figure}

In reality the number of UHECR events is of course finite,
and often not very large. 
The discriminatory power then depends not only
on the flux map but also on event number.
In this case,
equation~\eqref{eq:DKS-lim} provides a figure-of-merit
that governs the asymptotic value of $D$.
This is illustrated in Fig.~\ref{fig:DKS-NeV}, where we show
the average value of $D$ and the 1-$\sigma$ bands as a
function of event number. When $N_{\rm ev}$ becomes very
large, $D$ approaches the asymptotic value 
computed with eq.~\eqref{eq:DKS-lim}: $0.22$ for the KKKST catalogue and
$0.17$ for the PSCz catalogue.

\section{Summary}
\label{sec:summary}
In this paper we have investigated how the choice of galaxy catalogue affects
UHECR model fluxes in a scenario where UHECR sources trace the distribution of matter in
the Universe. The differences between the three catalogues considered here,
the KKKST catalogue \citep{Kalashev:2007ph},
the XSCz catalogue\footnote{See \texttt{http://web.ipac.caltech.edu/staff/jarrett/XSCz/}},
and the PSCz  catalogue \citep{Saunders:2000af} are reasonably small. This is reassuring
because all catalogues are supposed to be sampling the same underlying
density field (barring biases induced by selection effects).

In section~\ref{section:KKKST} we have shown that there is good agreement between
model flux maps constructed with the KKKST and XSCz catalogues. The former
was used by us earlier in \pIsp \citep{Koers:2008ba} 
The agreement is especially good at energies below 60 EeV, the regime 
where we have confronted models with experimental data in \pI.
This comparison refutes the recent statement by 
\citet{Kashti:2009ui} that the KKKST catalogue is not suited for studies of UHECR anisotropy
because of its dependence on photometric redshift estimates.

We have investigated the relation between
the predicted strength of large-scale UHECR anisotropies and model flux maps 
from a general point of view in section \ref{section:fluxmapcontrasts}.
The intrinsic properties of an UHECR flux map, i.e. those properties relating to
the strength of over- and underdense regions but not to their position on the sky,
are contained in the flux distribution. Equation \eqref{eq:DKS-lim} demonstrates
how this distribution can be used to determine the value of the $D$ test statistic
in the limit of infinite events. This asymptotic value provides a measure of the
expected anisotropy in UHECR arrival directions for sources tracing the
distribution of matter in the Universe. We have compared these values
for the KKKST, XSCz, and PSCz catalogues
as a function of energy (see Fig.~\ref{fig:DKS-E}). The comparison
shows that the KKKST catalogue typically yields stronger anisotropies than
the PSCz catalogue, the XSCz catalogue being in between. The difference is
however small in the light of the uncertainties induced by
systematical errors in UHECR energy determination.

\section*{acknowledgments}

We would like to thank Tom Jarrett for providing us
with a preliminary version of the XSCz catalogue, and Tamar Kashti and Sergey Troitsky
for stimulating discussions.
H.K. and P.T. are supported by Belgian Science Policy under IUAP VI/11
and by IISN. The work of P.T. is supported in part by the
FNRS, contract 1.5.335.08.


\bibliographystyle{mn2e}
\bibliography{refs}

\appendix
\section{Kolmogorov-Smirnov distance in the limit of infinite statistics}
\label{app:DKS}
In this appendix we derive equation \eqref{eq:DKS-lim}. We choose
a general setup because the result is not limited to UHECR anisotropy
tests, but is applicable to any experimental test that
can be appropriately formulated.

Consider an experiment that records events which are characterized
by a set of quantities $\vec{x}$ (these quantities may be observables
or derived quantities). The space  of
all possible experimental results is denoted as $V$.
Now consider two models, termed model $\mathbf{A}$ and model $\mathbf{B}$.
Model $\mathbf{A}$ asserts that events sample
$V$ uniformly, i.e. that the
probability of registering quantities in the range
$\vec{x} \ldots \vec{x} + d \vec{x}$
is independent of $\vec{x}$ (as long as $\vec{x} \in V$).
Within model $\mathbf{B}$, on the other hand,
the probability that an event has quantities in the range
$\vec{x} \ldots \vec{x} + d \vec{x}$ is proportional to
$f(\vec{x}) d \vec{x}$, where $f$ is a non-trivial, known function.
This function defines a map $\vec{x} \mapsto f$, which associates
a real number $f$ to every event. Note that we leave the normalization
of $f$ arbitrary.

Our aim is now to differentiate between models $\mathbf{A}$
and $\mathbf{B}$. 
We will do this by considering the
distribution of $f$ over an observation,
i.e. a series of registered events.
We consider the limit of infinite statistics.
The distribution of $f$ under model $\mathbf{A}$, denoted as
$\mathcal{D}_{\mathbf{A}} (f)$, represents the distribution of
$f$ over $V$.  We assume this distribution is normalized, i.e.
\beqs
\int_0^{\infty} df \mathcal{D}_{\mathbf{A}} (f) = 1 \, .
\eeqs
The distribution of $f$ under model $\mathbf{B}$ is denoted as
$\mathcal{D}_{\mathbf{B}} (f)$. By construction,
\beqs
\mathcal{D}_{\mathbf{B}} (f) = c f \mathcal{D}_{\mathbf{A}} (f) \, ,
\eeqs
where 
\beqs
c = \lb \int_0^{\infty} df f \mathcal{D}_{\mathbf{A}} (f) \rb^{-1} 
\eeqs
is a normalization constant to ensure that $\mathcal{D}_{\mathbf{B}} (f)$
is also normalized. Note that $1/c$ coincides with $\bar{f}_{\mathbf{A}}$,
the average value of $f$ in model $\mathbf{A}$.
We define cumulative distribution functions as
\beqs
\mathcal{C} (f)  = \int_0^f d f' \mathcal{D} (f') \, . 
\eeqs
The Kolmogorov-Smirnov distance $D_{\rm KS}$ 
is a measure of the difference between
$\mathcal{C}_{\mathbf{A}}$ and $\mathcal{C}_{\mathbf{B}}$.
As we will show, it can be expressed in terms
of the cumulative distribution function $\mathcal{C}_{\mathbf{A}} (f)$ alone.
First we recall that, by definition,
$ D_{\rm KS} = \max |D(f)|$,
where $D(f) = \mathcal{C}_{\mathbf{A}} (f) - \mathcal{C}_{\mathbf{B}} (f)$ denotes the
difference between the two cumulative distribution functions. This difference
can be expressed as follows:
\beqs
D(f)  = \int_0^f df' \mathcal{D}_{\mathbf{A}} (f') \lb 1- c f' \rb \, .
\eeqs
Since $\mathcal{D}_{\mathbf{A}}$, $c$ and $f'$ are strictly positive,
the integral is maximal when
$f=1/c = \bar{f}_{\mathbf{A}}$. Hence
\beqs
\label{eq:DKS:final}
D_{\rm KS} = D(\bar{f}_{\mathbf{A}})
= \frac{1}{\bar{f}_{\mathbf{A}}} \int_0^{\bar{f}_{\mathbf{A}}} d f' \mathcal{C}_{\rm A} (f') \, ,
\eeqs
where the last equality has been obtained by integration by parts.
Normalizing the function $f$ such that
$\bar{f}_{\mathbf{A}} = 1$, 
the r.h.s. of eq.~\eqref{eq:DKS:final} reduces to
the integral of the cumulative distribution function
$\mathcal{C}_{\mathbf{A}}$ up to $f=1$.

Two remarks are in order. First, equation~\eqref{eq:DKS:final} is a general result that
is applicable to any statistical test for a model that
can be formulated in terms of a probability function $f(\vec{x})$,
where $\vec{x}$ is distributed uniformly in $V$ under the alternative model.
The result applies to the UHECR anisotropy tests considered in this
work  by associating $\vec{x} = (l,\sin b)$, where $l$ and $b$ denote
galactic coordinates, and $f = \phi / \langle \phi \rangle$,
where $\phi$ denotes the model flux when UHECR sources
trace the distribution of matter, and $\langle \phi \rangle$
is the average value on the sphere. 
Note that, by definition, $\bar{f}_{\mathbf{A}} = 1$ in this case.
Second, equation~\eqref{eq:DKS:final}
determines $D_{\rm KS}$ in the limit of an infinite number of events. 
For finite event numbers, it is still a useful figure of merit because
it determines  the asymptotic behaviour (cf.~Fig.~\ref{fig:DKS-NeV}).

\label{lastpage}

\end{document}